\documentclass[12pt]{article}

\usepackage{amsmath,amsthm, amsfonts, amssymb, amsxtra, amsopn}
\usepackage{pgfplots}
\pgfplotsset{compat=1.13}
\usepackage{pgfplotstable}
\usepackage{graphicx,grffile}
\usepackage{multirow}
\usepackage{booktabs}
\usepackage{listings}
\usepackage{cmap}
\usepackage{colortbl}
\usepackage{adjustbox}
\usepackage{epsfig}

\usepackage[tableposition=top,font=small,skip=5pt]{caption}
\usepackage{subcaption}
\usepackage{makecell} 

\usepackage[explicit]{titlesec}

\PassOptionsToPackage{hyphens}{url}

\usepackage{hyperref}
\hypersetup{colorlinks=true,linkcolor=black,citecolor=black,urlcolor=blue,filecolor=black}
\hypersetup{pdfpagemode=UseNone,pdfstartview=}

\definecolor{darkgreen}{rgb}{0.125,0.5,0.169}

\usepackage{enumitem}
\setlist[itemize]{noitemsep, topsep=0pt}

\advance\oddsidemargin by -0.35in
\advance\textwidth by 0.7in

\advance\topmargin by -0.4in
\advance\textheight by 0.8in

\long\def\symbolfootnotetext[#1]#2{\begingroup%
\def\thefootnote{\fnsymbol{footnote}}\footnotetext[#1]{#2}\endgroup}

\newcommand\dunderline[3][-1pt]{{%
  \sbox0{#3}%
  \ooalign{\copy0\cr\rule[\dimexpr#1-#2\relax]{\wd0}{#2}}}}
\def\uuu{\kern-1pt\dunderline{0.75pt}{\phantom{M}}}

\DeclareMathOperator{\thth}{th}

\DeclareMathOperator{\rotimes}{{\color{red}\boldsymbol{\otimes}}}
\DeclareMathOperator{\roCat}{%
    \raisebox{-4pt}{%
    \begin{tikzpicture}%
        \draw[thick,color=blue] (0.0,0.0) circle(0.15);%
        \node at (0.0,0.0) {$\color{blue}\scriptstyle\boldsymbol{\|}$};%
    \end{tikzpicture}%
    }%
}

\title{Multimodal Techniques for Malware Classification}

\author{Jonathan Jiang\footnotemark[1]\ \ \ 
Mark Stamp\footnotemark[1]\,\,\footnotemark[2]}

\begin{document}

\symbolfootnotetext[1]{Department of Computer Science, San Jose State University}
\symbolfootnotetext[2]{mark.stamp$@$sjsu.edu}

\maketitle

\abstract
The threat of malware is a serious concern for computer networks and systems,
highlighting the need for accurate classification techniques. 
In this research, 
we experiment with multimodal machine learning approaches for malware classification,
based on the structured nature of the Windows Portable Executable (PE) file format. 
Specifically, we train 
Support Vector Machine (SVM), 
Long Short-Term Memory (LSTM), and 
Convolutional Neural Network (CNN) models on features extracted from PE headers, 
we train these same models on features extracted from the other sections of PE files, and 
train each model on features extracted from the entire PE file.
We then train SVM models on each of the nine header-sections combinations of these baseline models,
using the output layer probabilities of the component models as feature vectors. 
We compare the baseline cases to these multimodal combinations.
In our experiments, we find that the best of the multimodal models outperforms 
the best of the baseline cases, indicating that it can be advantageous to train separate 
models on distinct parts of Windows PE files.

\section{Introduction}

Rapid development in technology has changed the way humans use and acquire knowledge and interact with society, 
leading to an increased reliance on personal computers and smart devices. In recent decades,
valuable information has become available on the Internet and on personal computers, and such data 
is a natural target for cybercriminals~\cite{Alenezi2020EvolutionOM}. 
Cybercrime and malware go hand-in-hand, and hence malware detection and classification
are vital concerns in cybersecurity.

Windows Portable Executable (PE) files are the most common victims of malware attacks. 
Due to their popularity, malware creators often 
choose to attack PE file formats~\cite{PEFILEs}.

Traditional signature-based detection methods are fast, efficient, and accurate for known threats.
However, due to their dependency on predefined patterns or signatures of known malware, 
they lack adaptability and are vulnerable to evasion techniques. Malware that can change its code 
structure while maintaining malicious functionality can easily escape from the signature-based antivirus 
scanning~\cite{Machine_Learning_for_Windows}. As a result, machine learning and deep learning
techniques have become mainstays in the malware detection field.

In this research, we apply machine learning techniques to the problem of Windows PE malware classification
using a multimodal approach. In this context, multimodal means that we apply machine learning techniques 
that combine multiple types of data, with the goal of improving the accuracy of the system. 
Specifically, in this paper, we train learning models on different parts of PE files, then train another
machine learning model with the probability vectors of the component models serving as features.
While copious previous research has trained trained a wide variety learning models on a wide range of features, 
(byte sequences, opcode sequences, API calls, various graph structures, features derived from code analysis, metadata,
and so on, there is relatively little research that attempts to directly utilize the structure inherent in PE 
files~\cite{10049419,9844516,YUAN2020101740}.

This research expands upon traditional methods by focusing on the structural characteristics of PE files, 
which are composed of headers and various additional sections (heretofore referred to simply as sections) . 
We extract features from PE headers and from the PE sections. Then we train models just using the
header-based features, we train models just using the sections-based features, and we train models using
features from both the header and sections (i.e., the entire PE file). We then train multimodal learning models 
using the output of the header and section models as the feature vectors. In this context, the output of a model
refers to the probability vector generated by the output layer of the model. We then compare 
the results of these different cases, namely header-based models, section-based models,
models trained on the entire PE file, and multimodal models trained on the output of 
header and sections based models. Note that in the multimodal cases, the training of the
header based model and the sections based model can be viewed as a feature engineering
step, with the output of these models yielding additional features that may be more informative,
as compared to the original features used to train the component models.

For the header-based, sections-based models, and the models trained on the entire PE file,
we experiment with  
Support Vector Machines (SVM), 
Long Short-Term Memory (LSTM) models, and 
Convolutional Neural Networks (CNN). 
For the multimodal models we use SVMs to combine the output of the 
header-based and sections-based models.

The remainder of this paper is organized as follows. 
Section~\ref{sect:back} introduces relevant background topics, 
including an overview of the PE file format and the learning models employed in our experiments. 
In Section~\ref{sect:related} we discuss representative examples of relevant related work.
Section~\ref{sect:exp} presents our experiment, including dataset preparation, 
feature extraction, model design, hyperparameter tuning, and so on. In this section,
we also compare and analyze our results.  Section~\ref{sect:con} summarizes the key 
findings of this research and provides some suggestions for further work.

\section{Background}\label{sect:back}

The Portable Executable (PE) file format is the standard for Windows executables. 
Due to their ubiquity, PE files are the predominant target for malware creators. 
According to a~2020 report from Kaspersky, nearly~90\%\ of 
all malware detected daily occurred in PE files~\cite{kasp}. Given the frequency of this format in malware attacks, 
understanding the structure and functionality of PE files may be advantageous for malware analysis. 
Furthermore, a deep understanding of PE file structure facilitates the extraction of useful features 
for machine learning-based malware detection. Next, we provide an overview of the PE file format,
which is followed by an introduction to the three types of machine learning models considered
in this research.

\subsection{Overview of the PE File Format}

The PE file format is derived from the Microsoft Common Object File Format (COFF). 
It is a highly modular and structured file format that ensures portability and compatibility 
on various Windows platforms~\cite{PE_Format}. Compared to a single contiguous memory-mapped file, 
PE files utilize header information that provides detailed instructions to the Windows dynamic linker, 
enabling it to load code (including required libraries) into designated memory locations with 
the appropriate permissions. This approach is designed to ensure performance and safety during 
execution. Common PE format executable extensions include \texttt{.exe}, \texttt{.dll}, \texttt{.scr}, and \texttt{.sys}. 
While all of these file types adhere to the PE structure, each type include unique metadata or characteristics 
to support their specific use cases~\cite{PEFILE_OVERVIEW}.

\begin{figure}[!htb]
    \centering
    \includegraphics[scale=0.5]{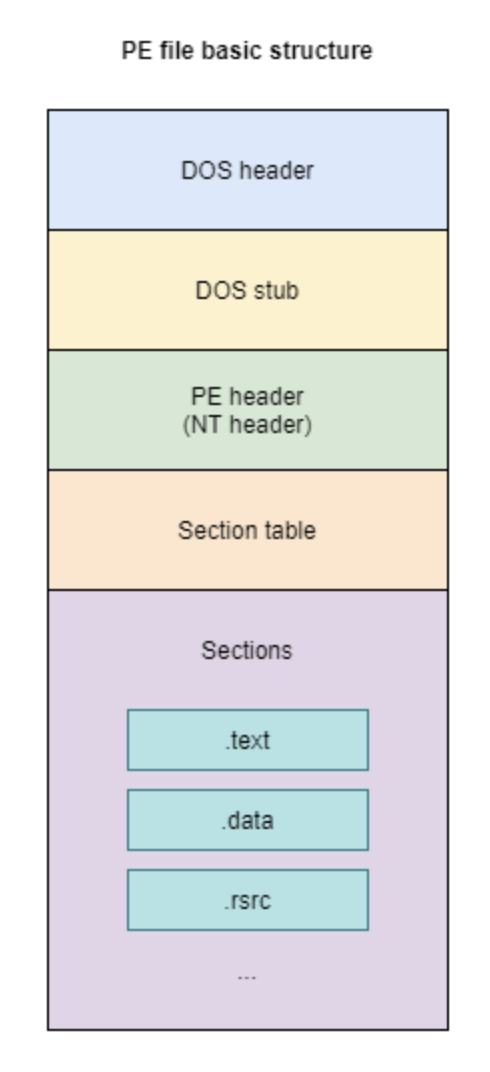}
    \caption{Structure of a Portable Executable (PE) file~\cite{PE_IMAGE}}
    \label{fig:pe_file_structure}
\end{figure}

An example layout of the Windows PE file is shown in Figure~\ref{fig:pe_file_structure}. 
Note that the PE file includes the DOS Header, DOS Stub, PE File Header (aka NT header), Section Table, 
and Sections. Next, we briefly discuss each of these components of a PE file.   

\begin{description}
\item[DOS Header]--- The DOS Header is the first data structure in a PE file. It provides backward compatibility 
with MS-DOS applications, which were applications designed to run in the Microsoft Disk Operating System (MS-DOS) environment during the beginning of 1980. The DOS Header occupies the first 64 bytes of the file, followed by a DOS stub program that outputs a message \texttt{"This program cannot be run in DOS mode"}. This message is used to notify the users that the file is currently executing in a DOS environment.
The first and last elements of the DOS Header are~\texttt{e\_magic} and~\texttt{e\_lfanew}, respectively.
This first element, \texttt{e\_magic}, is the two-byte ``magic number'' \texttt{0x5A4D} (or \texttt{MZ} in ASCII), 
which is used to simply denote the file as an MS-DOS executable.
The last member of the DOS Header structure, \texttt{e\_lfanew},
is the offset to the~NT headers. This field is crucial for the PE loader on Windows systems 
to determine where to locate the~NT headers to begin loading the executable.
An example of a DOS header is illustrated in Figure~\ref{fig:DOS_Headers}.
\begin{figure}[!htb] 
    \centering 
    \includegraphics[scale=0.45]{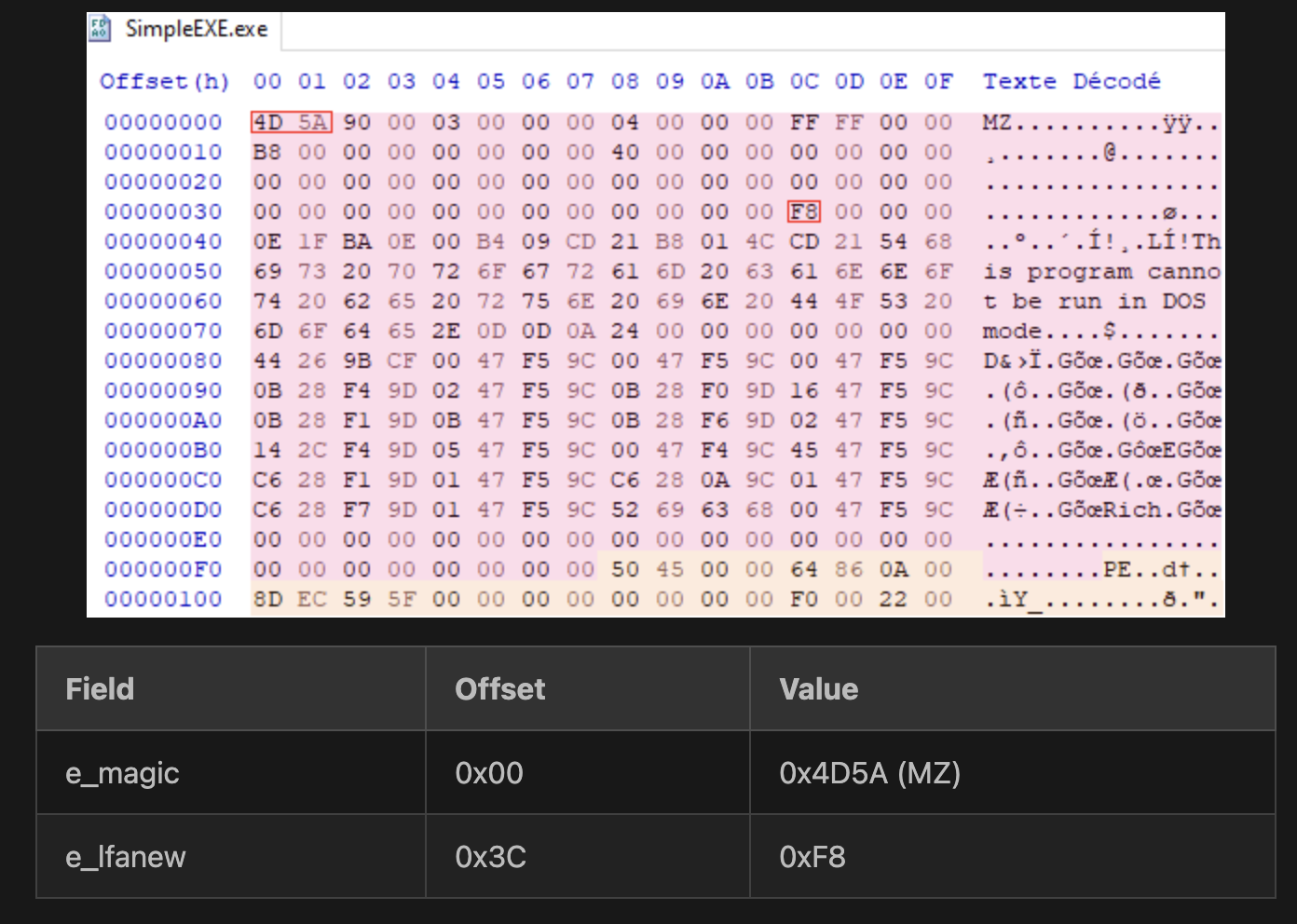}
    \caption{DOS header~\cite{portable_executable}}\label{fig:DOS_Headers}
\end{figure}
\item[PE Header]--- The PE Header (or NT header) begins with the PE signature (PE\textbackslash0\textbackslash0). 
It is a marker that identifies the file as a Portable Executable. It is followed by the COFF file header that 
provides metadata about the file, such as the target machine architecture, the number of sections,
the file creation timestamp, and attributes that describe the file's characteristics~\cite{PE_Format}.
\item[Optional Header]--- Contrary to its name, the Optional Header is not optional, as it is required for all PE image files. It provides necessary information like entry point address, the preferred memory location for loading, the size of the image in memory, and various attributes like subsystem type and security features. This header helps to guide the Windows loader in correctly loading and executing the program. However, it is optional for files like object files, which do not require runtime information~\cite{PE_Format}.
\item[Section Table]--- The Section Table is a series of section headers. Each entry in the table provides information for a specific section of the file, including the section name, virtual size, virtual address, raw data size, and assigned permissions. These headers contain the crucial information required by the loader to correctly map each section into memory with appropriate attributes, such as read, write, or execute permissions. Both the Section Table and the Optional Header play important roles in the loading and execution of a PE file but address different levels of detail: the Section Table focuses on sections, while the Optional Header provides a high-level view of the entire file and execution requirements.
\item[Sections]--- The Sections represent the actual contents of the PE file, including executable code, initialized and uninitialized data, and various embedded resources. Common sections typically found in most PE files include the following.
\begin{description}
\item[\texttt{.text}:]  The .text section contains the executable instructions for the application. Within this section, the entry point can be found. In this research, it served as the starting point for section-based feature extraction. Notably, An application may include multiple sections containing executable instructions.
\item[\texttt{.data}:] The .data section holds initialized data, such as global and static variables. It often includes constants and strings defined in the source code.
\item[\texttt{.rdata} \normalfont or \texttt{.idata}:] The .radata or .idata sections generally store read-only data and the import table. The import table lists the Windows API functions used by the executable file, along with the names of their associated dynamic link libraries. This information allows the Windows loader to locate and link the required API functions from the appropriate system dynamic link libraries.
\item[\texttt{.reloc}:] The .reloc session contains the relocation table, which provides information needed when the executable is loaded at a base address different from the one initially specified. It helps adjust the addresses of code and data references accordingly.
By default, most executable files assume a preferred base address. It will be the starting address in memory where they expect to be loaded. If the program is loaded at the preferred base address, then all its hard-coded addresses will be correct, and the program can run without modification. However, if the executable is loaded to a different base address, this section would provide the necessary information. 
\item[\texttt{.rsrc}:] The .rsrc section contains various embedded resources used by the executable, such as icons, dialogs, images, menus, and version information. These resources are typically part of the application’s user interface.
\item[\texttt{.debug}:] This .debug section stores debugging information, including symbols and source line references. It is used by debuggers to provide meaningful diagnostics during development.
\end{description}
\end{description}

\subsection{Machine Learning Methods}\label{sect:ML_methods}

This section introduces the machine learning that are used in the experiments considered in this paper. 
Specifically, we discuss Support Vector Machines (SVM), Long Short-Term Memory (LSTM) models, 
and Convolutional Neural Networks (CNN).

\subsubsection{Support Vector Machines}

The main ideas behind SVMs are the following.
\begin{itemize}
\item Separating hyperplane --- The goal is to separate labeled training data 
into two classes based on a hyperplane.
\item Maximize the margin --- When constructing the separating hyperplane, 
we maximize the margin,
which is defined as the minimum separation between the two classes
in the training set.
\item Work in a higher dimensional space --- By moving the 
problem to a higher dimension, there is more space
available, and hence there is a better chance of finding
a separating hyperplane. 
\item Kernel trick --- A kernel function can be used to transform the data 
to a higher dimensional space, with the goal of obtaining better separation.
The reason that this is considered a ``trick'' is because we are able to
work in a higher dimensional space, while paying a minimal
performance penalty. 
\end{itemize}

\begin{figure}[!htb]
    \centering
        \begin{tikzpicture}[scale=0.925]
    
    \draw[thick,color=blue] (4,2) rectangle (4.15,2.15);
    \draw[thick,color=blue] (3.5,4.25) rectangle (3.65,4.4);
    \draw[thick,color=blue] (3.2,2.0) rectangle (3.35,2.15);
    \draw[thick,color=blue] (3.0,2.75) rectangle (3.15,2.9);
    \draw[thick,color=blue] (3.45,2.65) rectangle (3.6,2.8);
    \draw[thick,color=blue] (3.75,2.7) rectangle (3.9,2.85);
    \draw[thick,color=blue] (3.5,3.25) rectangle (3.65,3.4);
    \draw[thick,color=blue] (3.0,3.5) rectangle (3.15,3.65);
    \draw[thick,color=blue] (2,3) rectangle (2.15,3.15);
    \draw[thick,color=blue] (2.5,3.5) rectangle (2.65,3.65);
    \draw[thick,color=blue] (2.25,4.35) rectangle (2.4,4.5);
    \draw[thick,color=blue] (3.6,1.5) rectangle (3.75,1.65);
    
    \draw[thick,color=red,fill=red] (1.4,1.2) circle (0.08);
    \draw[thick,color=red,fill=red] (0.6,0.75) circle (0.08);
    \draw[thick,color=red,fill=red] (0.95,0.5) circle (0.08);
    \draw[thick,color=red,fill=red] (1.925,0.925) circle (0.08);
    \draw[thick,color=red,fill=red] (0.5,1.25) circle (0.08);
    \draw[thick,color=red,fill=red] (1.1,1.0) circle (0.08);
    \draw[thick,color=red,fill=red] (2.55,0.25) circle (0.08);
    \draw[thick,color=red,fill=red] (1.05,1.5) circle (0.08);
    \draw[thick,color=red,fill=red] (0.9,1.825) circle (0.08);
    \draw[thick,color=red,fill=red] (0.5,1.75) circle (0.08);
    

    
    \draw[thick,color=brown] (0,4) -- (4,0); 
    \draw[thick,dashed,color=blue] (0,5) -- (5,0); 
    \draw[thick,dashed,color=red] (0,3) -- (3,0); 

    \draw[thick,color=black,<->] (2,1) -- (2.5,1.5); 
    \draw[thick,color=black,<->] (1.5,2.5) -- (2,3); 
    
     \draw[thick,color=black,->] (0,0) -- (6,0); 
     \draw[thick,color=black,->] (0,0) -- (0,5); 
     \draw[thick,color=white,->] (0,-0.2) -- (6,-0.2); 
     \draw[thick,color=white,->] (-0.2,0) -- (-0.2,5); 

    \end{tikzpicture}
    \caption{The separating hyperplane of an SVM}\label{fig:SVM_1}
\end{figure}
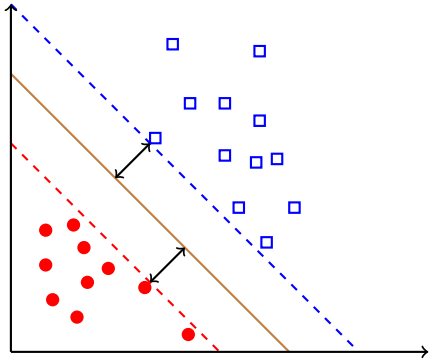

Figure~\ref{fig:SVM_1} provides a generic view of linearly separable data, 
along with the separating hyperplane of an SVM, which maximizes the margin.

According to~\cite[p.~9]{BenCamp}, 
``SVMs are a rare example of a methodology where geometric
intuition, elegant mathematics, theoretical guarantees, and
practical algorithms meet.''
We note in passing that the SVM technique 
easily generalizes to multiclass data,
in which case the technique is sometimes referred to as 
Support Vector Classifier (SVC).

\subsubsection{Long Short-Term Memory}\label{sect:LSTM}

In this section, we
present long short-term memory (LSTM)
networks in some detail.
A vast number of variants of the LSTM architecture have been developed.
However, according to an extensive empirical study~\cite{GreffSKSS17}, 
``none of the variants can improve upon the standard 
LSTM architecture significantly.''

In addition to being a tongue twister,
LSTM networks are a class of RNN 
architectures that are designed to deal with long-range dependencies.
That is, LSTM can deal with ``gaps'' between the appearance of a feature and the point
at which it is used by the model~\cite{GreffSKSS17}. The claim to fame of LSTM is
that it can mitigate the effect of vanishing gradients, which is what enables 
such models to account for longer-range dependencies~\cite{SH_JS}.

Before outlining the ideas behind LSTM, we
note that the LSTM architecture has been one of the most 
commercially successful learning techniques ever developed. 
Among many other applications, LSTMs have
been used in \hbox{Google} \hbox{Allo}~\cite{allo}, 
\hbox{Google} \hbox{Translate}~\cite{translate}, 
\hbox{Apple's} \hbox{Siri}~\cite{iBrain},
and \hbox{Amazon} \hbox{Alexa}~\cite{alexa}. 
However, the commercial dominance of
LSTM may be waning~\cite{explode}.

Figure~\ref{fig:LSTM} illustrates a generic LSTM. From this high level
perspective, an obvious difference from a
plain vanilla RNN is that an LSTM has two ``lines'' entering and 
exiting each state. As in a standard RNN, one of these lines represents the hidden state,
while the second line is designed to
serve as a gradient ``highway'' during backpropagation. In this way,
the gradient can ``flow'' much further back with less chance that it will 
vanish along the way.

\begin{figure}[!htb]
\centering
  \begin{tikzpicture}[scale=0.75,every node/.style={scale=0.85}]

    \draw[thick,color=green] (-2.0,5.5) circle (0.575);
    \draw[thick,color=blue] (-2.5,2.5) rectangle (-1.5,3.5);
    \draw[thick,color=green,->] (-2.0,4.925) -- (-2.0,3.5);
    \draw[thick,color=blue,->] (-2.0,2.5) -- (-2.0,1.17);
    \draw[thick,color=blue,->] (-1.5,2.75) -- (0.0,2.75);
    \draw[thick,color=blue,->] (-1.5,3.25) -- (0.0,3.25);

    \node at (-2.0,5.5) {$X_{0}$};
    \node at (-2.0,3.0) {$L_{0}$};
    \node at (-2.0,0.75) {$h_{0}$};

    \draw[thick,color=green] (0.5,5.5) circle (0.575);
    \draw[thick,color=blue] (0.0,2.5) rectangle (1.0,3.5);
    \draw[thick,color=green,->] (0.5,4.925) -- (0.5,3.5);
    \draw[thick,color=blue,->] (0.5,2.5) -- (0.5,1.17);
    \draw[thick,color=blue,->] (1.0,2.75) -- (2.5,2.75);
    \draw[thick,color=blue,->] (1.0,3.25) -- (2.5,3.25);

    \node at (0.5,5.5) {$X_{1}$};
    \node at (0.5,3.0) {$L_{1}$};
    \node at (0.5,0.75) {$h_{1}$};
    
    \draw[thick,color=green] (3.0,5.5) circle (0.575);
    \draw[thick,color=blue] (2.5,2.5) rectangle (3.5,3.5);
    \draw[thick,color=green,->] (3.0,4.925) -- (3.0,3.5);
    \draw[thick,color=blue,->] (3.0,2.5) -- (3.0,1.17);
    \draw[thick,color=blue,->] (3.5,2.75) -- (5.0,2.75);
    \draw[thick,color=blue,->] (3.5,3.25) -- (5.0,3.25);

    \node at (3.0,5.5) {$X_{2}$};
    \node at (3.0,3.0) {$L_{2}$};
    \node at (3.0,0.75) {$h_{2}$};
    
    \draw[thick,color=green] (5.5,5.5) circle (0.575);
    \draw[thick,color=blue] (5.0,2.5) rectangle (6.0,3.5);
    \draw[thick,color=green,->] (5.5,4.925) -- (5.5,3.5);
    \draw[thick,color=blue,->] (5.5,2.5) -- (5.5,1.17);
    \draw[thick,color=blue,->] (6.0,2.75) -- (7.5,2.75);
    \draw[thick,color=blue,->] (6.0,3.25) -- (7.5,3.25);

    \node at (5.5,5.5) {$X_{3}$};
    \node at (5.5,3.0) {$L_{3}$};
    \node at (5.5,0.75) {$h_{3}$};


    \node[color=green] at (8.0,5.5) {\large$\cdots$};
    \node[color=blue] at (8.0,2.75) {\large$\cdots$};
    \node[color=blue] at (8.0,3.25) {\large$\cdots$};
    \node[color=black] at (8.0,0.75) {\large$\cdots$};
    
\end{tikzpicture}
\caption{LSTM}\label{fig:LSTM}
\end{figure}

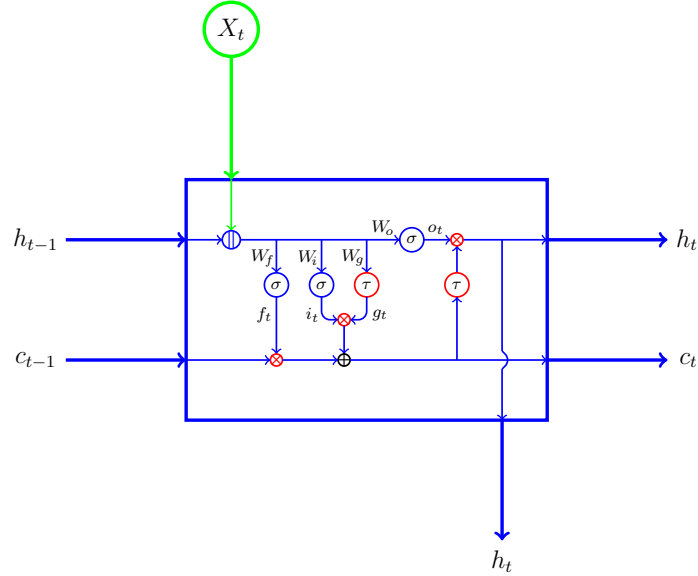
\begin{figure}[!htb]
\centering
\adjustbox{scale=0.80}{
  \begin{tikzpicture}[scale=1.0,every node/.style={scale=1.0}]

    \draw[ultra thick,color=blue] (0.0,0.0) rectangle (6.0,4.0);
    \draw[ultra thick,color=green] (0.75,6.5) circle(0.45);
    \draw[ultra thick,color=green,->] (0.75,6.05) -- (0.75,4.0);
    \draw[ultra thick,color=blue,->] (5.25,0.0) -- (5.25,-2.0);
    \draw[ultra thick,color=blue,->] (-2.0,1.0) -- (0.0,1.0);
    \draw[ultra thick,color=blue,->] (-2.0,3.0) -- (0.0,3.0);
    \draw[ultra thick,color=blue,->] (6.0,1.0) -- (8.0,1.0);
    \draw[ultra thick,color=blue,->] (6.0,3.0) -- (8.0,3.0);

    \node at (-2.5,1.0) {$c_{t-1}$};
    \node at (-2.5,3.0) {$h_{t-1}$};
    \node at (0.75,6.5) {$X_{t}$};
    \node at (5.25,-2.35) {$h_{t}$};
    \node at (8.35,1.0) {$c_{t}$};
    \node at (8.325,3.0) {$h_{t}$};


    \draw[thick,color=green,->] (0.75,4.0) -- (0.75,3.15);
    \draw[thick,color=blue] (0.75,3.0) circle(0.15);
    \node at (0.75,3.0) {$\color{blue}\scriptstyle\boldsymbol{\|}$};

    \draw[thick,color=blue,->] (0.0,3.0) -- (0.60,3.0);
    \draw[thick,color=blue,->] (0.90,3.0) -- (3.55,3.0);
    \draw[thick,color=blue] (3.75,3.0) circle(0.2);
    \node at (3.75,3.0) {$\scriptstyle\sigma$};
    \node at (3.27,3.2) {$\scriptstyle W_{\kern -1.5pt o}$};
    \draw[thick,color=blue,->] (3.95,3.0) -- (4.4,3.0);
    \node at (4.15,3.175) {$\scriptstyle o_t$};


    \draw[thick,color=blue,->] (1.5,3.0) -- (1.5,2.45);
    \draw[thick,color=blue] (1.5,2.25) circle(0.2);
    \node at (1.5,2.25) {$\scriptstyle\sigma$};
    \node at (1.275,2.7) {$\scriptstyle W_{\kern -1.5pt f}$};
    \draw[thick,color=blue,->] (1.5,2.05) -- (1.5,1.1);
    \node at (1.3,1.75) {$\scriptstyle f_t$};
    \draw[thick,color=red] (1.5,1.0) circle(0.1);
    \node at (1.5,1.0) {$\color{red}\scriptstyle\boldsymbol{\times}$};         
 
    \draw[thick,color=blue,->] (2.25,3.0) -- (2.25,2.45);
    \draw[thick,color=blue] (2.25,2.25) circle(0.2);
    \node at (2.025,2.7) {$\scriptstyle W_{\kern -1.5pt i}$};
    \node at (2.25,2.25) {$\scriptstyle\sigma$};    
    \draw[thick,color=blue,rounded corners,->] (2.25,2.05) -- (2.25,1.67) -- (2.525,1.67);
    \node at (2.1,1.75) {$\scriptstyle i_t$};

    \draw[thick,color=blue,->] (3.0,3.0) -- (3.0,2.45);
    \node at (2.775,2.7) {$\scriptstyle W_{\kern -1.5pt g}$};
    \draw[thick,color=red] (3.0,2.25) circle(0.2);
    \node at (3.0,2.25) {$\scriptstyle\tau$};     
    \draw[thick,color=blue,rounded corners,->] (3.0,2.05) -- (3.0,1.67) -- (2.725,1.67);
    \draw[thick,color=red] (2.625,1.67) circle(0.1);
    \node at (2.625,1.67) {$\color{red}\scriptstyle\boldsymbol{\times}$};         
    \draw[thick,color=blue,->] (2.625,1.57) -- (2.625,1.1);
    \draw[thick,color=black] (2.625,1.0) circle(0.1);
    \node at (2.625,1.0) {$\scriptstyle\hbox{}\boldsymbol{+}\hbox{}$};         
    \node at (3.225,1.75) {$\scriptstyle g_t$};


    \draw[thick,color=blue,->] (4.5,2.45) -- (4.5,2.9);
    \draw[thick,color=red] (4.5,2.25) circle(0.2);
    \node at (4.5,2.25) {$\scriptstyle\tau$};    
    \draw[thick,color=blue,->] (4.5,1.0) -- (4.5,2.05);

    \draw[thick,color=blue,->] (4.6,3.0) -- (6.0,3.0);
    \draw[thick,color=red] (4.5,3.0) circle(0.1);
    \node at (4.5,3.0) {$\color{red}\scriptstyle\boldsymbol{\times}$};         

    \draw[thick,color=blue,->] (0.0,1.0) -- (1.4,1.0);
    \draw[thick,color=blue,->] (1.6,1.0) -- (2.525,1.0);
    \draw[thick,color=blue,->] (2.725,1.0) -- (6.0,1.0);

    \draw[thick,color=blue] (5.25,3.0) -- (5.25,1.15);
    \draw[thick,color=blue,->] (5.25,0.85) -- (5.25,0.0);
    \draw[thick,color=blue,rounded corners] (5.25,1.15) -- (5.4,1.0) -- (5.25,0.85);
    
\end{tikzpicture}
}
\caption{One timestep of an LSTM}\label{fig:LSTM_1}
\end{figure}

In Figure~\ref{fig:LSTM_1} we expand one LSTM cell~$L_t$ 
that appears in Figure~\ref{fig:LSTM}.
Here, $\sigma$ is the sigmoid function, $\tau$ is the hyperbolic tangent (i.e., $\tanh$) function, 
the operators~``$\rotimes$'' and~``$\boldsymbol{\oplus}$'' are 
pointwise multiplication and addition, respectively,  while~``$\roCat$'' 
indicates concatenation of vectors.
The vector~$i_t$ is the ``input gate,'' $f_t$ is the ``forget gate,'' and~$o_t$ is the ``output gate.''
The vector~$g_t$ is an intermediate gate and 
is sometimes referred to as the ``gate gate''~\cite{RNN_Stanford}. 
We have much more to say about these gates below.

The gate vectors that appear in Figure~\ref{fig:LSTM_1} are computed as 
$$
\begin{array}{ll}
  f_t = \sigma\Biggl(W_{\kern -1.5pt f} 
		\biggl(\begin{array}{c}
			\kern-3pt h_{t-1}\kern-6pt \\
			\kern-3pt X_{t\phantom{-1}}\kern-6pt
		\end{array}\biggr)
 	+ b_f\Biggr) 
&
  g_t = \tau\Biggl(W_{\kern -1.5pt g} 
		\biggl(\begin{array}{c}
			\kern-3pt h_{t-1}\kern-6pt \\
			\kern-3pt X_{t\phantom{-1}}\kern-6pt
		\end{array}\biggr)
  	+ b_g\Biggr) \\ \\[-1ex]
  i_t = \sigma\Biggl(W_{\kern -1.5pt i} 
		\biggl(\begin{array}{c}
			\kern-3pt h_{t-1}\kern-6pt \\
			\kern-3pt X_{t\phantom{-1}}\kern-6pt
		\end{array}\biggr)
  	+ b_i\Biggr)
&
  o_t  = \sigma\Biggl(W_{\kern -1.5pt o} 
		\biggl(\begin{array}{c}
			\kern-3pt h_{t-1}\kern-6pt \\
			\kern-3pt X_{t\phantom{-1}}\kern-6pt 
		\end{array}\biggr)
  	+ b_o\Biggr)
\end{array}
$$
where the~$b$ terms are biases. The outputs are given by
\begin{align*}
  c_t &= f_t \rotimes c_{t-1} \boldsymbol{\oplus} i_t \rotimes g_t \\
  h_t &= o_t \rotimes \tau(c_t)
\end{align*}
where, as mentioned above, ``$\rotimes$'' is pointwise multiplication and~``$\boldsymbol{\oplus}$'' is the
usual pointwise addition. Note that each of the weight matrices is~$n\times 2n$.

In matrix form, ignoring the bias terms, we have
$$
  \left(\begin{array}{c}
  i_t \\
  f_t \\
  o_t \\
  g_t 
  \end{array}\right)
  =
  \left(\begin{array}{c}
  \sigma \\
  \sigma \\
  \sigma \\
  \tau 
  \end{array}\right)
  W
	\biggl(\begin{array}{c}
		\kern-3pt h_{t-1}\kern-6pt \\
		\kern-3pt X_{t\phantom{-1}}\kern-6pt 
	\end{array}\biggr)
$$
where~$X_{t}$ and~$h_{t-1}$ are column vectors of length~$n$,
and~$W$ is a~$4n\times 2n$ weight matrix of the form
$$
  W = \left(\begin{array}{c}
    W_{\kern -1.5pt i} \\
    W_{\kern -1.5pt f} \\
    W_{\kern -1.5pt o} \\
    W_{\kern -1.5pt g}
  \end{array}\right)\!.
$$
Each of the gates~$i_t$, $f_t$, $o_t$, and~$g_t$ is a
column vector of length~$n$. Recall that the sigmoid~$\sigma$ 
squashes its input to be within the range of~0 to~1, whereas the $\tanh$ 
function~$\tau$ produces output within the range of~$-1$ to~$+1$.

To highlight the intuition behind LSTM, we follow a similar
approach as that given in the excellent presentation at~\cite{RNN_Stanford}. 
Specifically, we focus on the extreme cases, that is, we pretend 
that the output of each sigmoid~$\sigma$ is either~0 or~1, and
each hyperbolic tangent~$\tau$ is either~$-1$ or~$+1$. Under this assumption, 
the forget gate~$f_t$ is a vector of~0s and~1s, where the~0s
tell us the elements of~$c_{t-1}$ that the model will forget (i.e., set to~0) and 
the~1s indicate the elements that the model will remember (i.e., retain). 

Again, assuming we only have the extreme cases, the input gate~$i_t$
and gate gate~$g_t$ together determine which elements of~$c_{t-1}$
to increment or decrement. Specifically, when element~$j$ of~$i_t$ 
is~1 and element~$j$ of~$g_t$ is~$+1$, we increment
element~$j$ of~$c_{t-1}$. On the other hand, if element~$j$ of~$i_t$ 
is~1 and element~$j$ of~$g_t$ is~$-1$, then we decrement
element~$j$ of~$c_{t-1}$. This serves to emphasize or de-emphasize
particular elements in the new-and-improved cell state~$c_t$.

Finally, the output gate~$o_t$ determines which elements of
the cell state will become part of the hidden state~$h_t$.
Note that the hidden state~$h_t$ is fed into the output layer of the LSTM.
Also note that before the cell states are operated on by the output
gate, the values are first squeezed down to be within 
the range of~$-1$ to~$+1$ by the~$\tau$ function.

Of course, in general, the LSTM gates are not simply counters that 
increment or decrement. But, the intuition holds, that is, the
gates keep track of incremental changes, thus allowing relevant information
to ``flow'' over long distances via the cell state. 
In this way, an LSTM can mitigate the 
gradient issues that arise in plain vanilla RNNs.

\subsubsection{Convolution Neural Network}

Artificial Neural Networks typically use fully connected layers, 
where a neuron at one layer is connected
to all neurons at the layers above and below.
A fully connected layer can deal effectively with correlations 
between any elements within the training vectors.
However, due to these fully connected layers,
the number of weight that must be determined
via training can be vast.
In contrast, a Convolution Neural Network (CNN), 
is designed to deal with local structure, which results in
a dramatic reduction in the number of weights that must be learned
during training. Thus, a key benefit of CNNs is that convolutional layers can be 
trained far more efficiently than fully connected layers.

For images, most of the relevant structure (edges and gradients, for example) 
is local. Hence, CNNs would seem to be an ideal tool for
image analysis.
However, CNNs have performed well in a wide variety of other problem domains.
In general, any problem for which there exists a data representation
where local structure predominates is a candidate for a CNN.
In addition to images, local structure is of primary importance in fields such as
text analysis and speech recognition, for example.

\begin{figure}[!htb]
    \centering
\begin{tikzpicture}[scale=0.4]

\draw[red,ultra thick] (0.0,0.0) rectangle (11.2,11.2);

\draw[green,ultra thick] (0.5,-0.5) rectangle (11.7,10.7);

\draw[blue,ultra thick] (1.0,-1.0) rectangle (12.2,10.2);


\draw[ultra thick, dotted] (0.0,9.1) rectangle (2.1,11.2);
\draw[ultra thick] (0.0,11.2) -- (2.1,11.2);
\draw[ultra thick] (0.0,11.2) -- (0.0,9.1);
\draw[ultra thick] (1.0,8.1) rectangle (3.1,10.2);
\draw[ultra thick] (0.0,9.1) -- (1.0,8.1);
\draw[ultra thick] (2.1,11.2) -- (3.1,10.2);
\draw[ultra thick] (0.0,11.2) -- (1.0,10.2);
\draw[ultra thick, dotted] (3.1,8.1) -- (2.1,9.1);

\draw[black,ultra thick] (13.6,1.4) rectangle (23.4,11.2);
\draw[ultra thick] (13.6,11.2) rectangle (14.3,10.5);
\draw[black,ultra thick] (14.1,0.9) rectangle (23.9,10.7);
\draw[ultra thick] (14.1,10.7) rectangle (14.8,10.0);
\draw[black,ultra thick] (14.6,0.4) rectangle (24.4,10.2);
\draw[ultra thick] (14.6,10.2) rectangle (15.3,9.5);
\draw[black,ultra thick] (15.1,-0.1) rectangle (24.9,9.7);
\draw[ultra thick] (15.1,9.7) rectangle (15.8,9.0);
\draw[black,ultra thick] (15.6,-0.6) rectangle (25.4,9.2);
\draw[ultra thick] (15.6,9.2) rectangle (16.3,8.5);

\draw[smooth,rounded corners,thick,->] (2.5,10.7) -- (2.5,11.9) -- (3.0,12.2) -- (12.95,12.2)  
	-- (13.55,11.9) -- (13.95,11.2);
\node at (7,12.7) {$\mbox{\footnotesize filter}_1$};
\draw[smooth,rounded corners,thick,->] (2.0,10.7) -- (2.0,12.9) -- (2.5,13.2) -- (12.95,13.2)  
	-- (13.55,12.9) -- (14.45,10.7);
\node at (7,13.7) {$\mbox{\footnotesize filter}_2$};
\draw[smooth,rounded corners,thick,->] (1.5,10.7) -- (1.5,13.9) -- (2.0,14.2) -- (12.95,14.2)  
	-- (13.55,13.9) -- (14.95,10.2);
\node at (7,14.7) {$\mbox{\footnotesize filter}_3$};
\draw[smooth,rounded corners,thick,->] (1.0,10.7) -- (1.0,14.9) -- (1.5,15.2) -- (12.95,15.2)  
	-- (13.55,14.9) -- (15.45,9.7);
\node at (7,15.7) {$\mbox{\footnotesize filter}_4$};
\draw[smooth,rounded corners,thick,->] (0.5,10.7) -- (0.5,15.9) -- (1.0,16.2) -- (12.95,16.2)  
	-- (13.55,15.9) -- (15.95,9.2);
\node at (7,16.7) {$\mbox{\footnotesize filter}_5$};

\end{tikzpicture}
    \caption{A convolutional layer}\label{fig:conv2}
\end{figure}
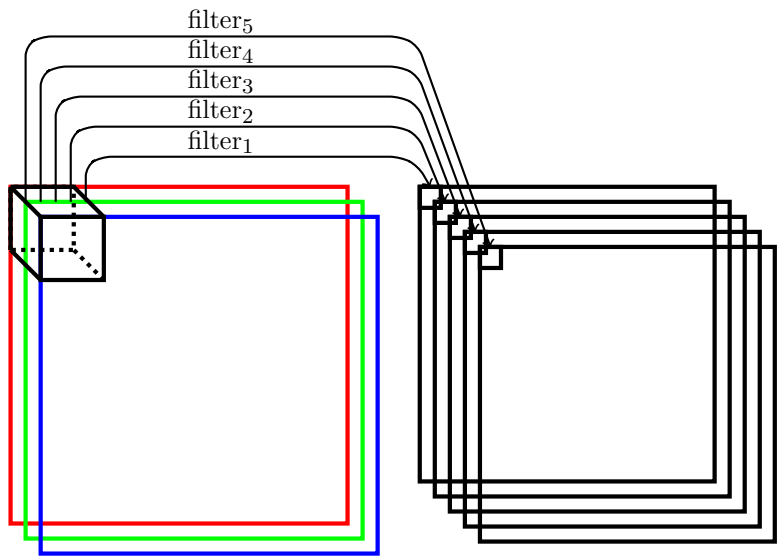

A convolutional layer is illustrated in Figure~\ref{fig:conv2}. In this case, 
five filters are trained on an RGB color image. By initializing the filters with different
values, they can potentially learn different features.

CNNs typically employ multiple convolutional layers. The first convolutional 
layer learns intuitive features, such as edges and gradients, while higher 
convolutional layers learn progressively more abstract features,
which ultimately allows for discrimination between complex classes, 
such as ``cat'' and ``dog''. Pooling layers are frequently 
employed between convolutional layers, which serve to reduce the dimensionality,
while also playing a role in translation invariance. Various regularization techniques
are used when training CNNs so as to mitigate issues related to overfitting.
For example, cutout regularization forces the model to consider different
parts of an image when training.

A detailed discussion of CNNs can be found at~\cite{Karpathy},
while the paper~\cite{Cornelisse} provides some interesting insights. For a more
intuitive discussion of CNNs, see~\cite{Kalfas}, and visual representations
can be found at~\cite{aditDeshpande}
For our purposes, it is important to note that
image-based analysis has recently proven highly effective 
for malware classification and analysis~\cite{Jain,Prajapati2021,Yajamanam}.

\section{Related Work}\label{sect:related}

In this section, we review examples of related work where PE file features 
have been used for malware classification. But first we mention that, in general, 
features used for malware analysis can be considered static or dynamic~\cite{Damodaran_2015}. 
Static features are those that can be obtained directly from the the malware file, whereas dynamic features
require code execution or emulation. Examples of static features include byte histograms and opcode
sequences, whereas API calls and various graph-based features typically must be collected
in a dynamic mode. Dynamic features are generally more robust with respect to obfuscation
techniques, while it is typically fare more efficient to collect static features. In the 
experiments reported in this paper, we only consider static features.

In~\cite{EMBER}, the EMBER dataset is introduced and analyzed. This dataset consists of a large set of static features extracted 
from~1.1 million PE files. The features include byte histograms, metadata in the header, and import table characteristics. 
The result from this study shows that machine learning techniques applied to these static features 
can distinguish between malware and benign with good accuracy.

Rezaei et al.~\cite{REZAEI2021102876} proposed a method focusing on using features from the PE header. 
They combined deep learning techniques and clustering techniques with~324 bytes of the header 
to train a malware classification model. The result of this model show good accuracy in malware classification, 
even when faced with obfuscation techniques. Similarly, Kattamuri et al.~\cite{electronics12020342} developed 
the SOMLAP dataset, which includes~51,409 samples, and from each sample, they extract~108 PE header attributes.
They first apply feature reduction techniques using swarm optimization algorithms,
including Ant Colony Optimization (ACO), Cuckoo Search Optimization (CSO), and Grey Wolf Optimization (GWO), 
which is followed by traditional machine-learning techniques. This research claimed to achieve a high malware detection 
accuracy. The two papers discussed in this paragraph illustrate the potential effectiveness of using header-based features.

Raff et al.~\cite{raff2017malwaredetectioneatingexe} proposed an end-to-end deep learning approach that avoids the need for 
manual feature selection. This approach feeds the entire byte sequence of PE files directly into a neural network,
skipping typical feature engineering steps. This allows the network to independently learn meaningful feature patterns. 
This approach showed high accuracy on large-scale datasets with one downside being heavy computational time.
In addition, the technique struggled with obfuscated malware.

Damodaran~\cite{Damodaran_2015} evaluated static, dynamic, and hybrid feature sets using Hidden Markov Models (HMMs). 
This paper highlights the trade-offs between different feature selection strategies and serves as a useful reference 
for optimizing feature selection methods. Static features like opcode sequences were found to be almost as effective 
as dynamic features in most cases. Hybrid approaches, which combine static and dynamic features, introduced 
additional complexity and failed to outperform standard approaches. 

Finally, Wen and Chow~\cite{WEN2021301128} proposed a malware detection model based on CNNs. 
In this research, variable-size binary fragments obtained from PE files are considered. The research simulated 
real-world scenarios where network packet size limitations result in incomplete file captures, aiming to demonstrate 
the ability to deal with such fragmented input. The dataset was derived from antivirus tools and included examples of
zero-day malware. Even with these challenges, the model still performed relatively well.

Table~\ref{tab:related_work} summarizes the related work discussed in this section.
Taken as a whole,
these studies demonstrate that a wide variety of PE-based feature selection and extraction methods have 
been used to build high-performing malware detection systems. Typical features include header fields, 
byte sequences, specific sections of the file, and even the full binary content analyzed using learning techniques. 
Together, these features enhance the performance of both traditional machine learning and modern deep learning 
models in malware classification.

\begin{table}[!htb]
\centering
\caption{Summary of related work in malware domain}\label{tab:related_work}
\adjustbox{scale=0.85}{
\begin{tabular}{l|l}
\toprule
\multicolumn{1}{c|}{\textbf{Research}\hspace*{0.35in}} & \multicolumn{1}{c}{\textbf{Main results}\hspace*{1.5in}} \\
\midrule
Anderson and Roth~\cite{EMBER} & Combining static features improved malware detection \\
\midrule
Rezaei, et al.~\cite{REZAEI2021102876} & Header-based features more robust against obfuscation \\
\midrule
Kattamuri, et al.~\cite{electronics12020342} & High accuracy with swarm optimization on header features \\
\midrule
\multirow{2}{*}{Raff, et al.~\cite{raff2017malwaredetectioneatingexe}} & Achieved high accuracy using PE file byte sequences, \\
                                                                                                              &  but struggled with obfuscation \\
\midrule
\multirow{2}{*}{Damodaran, et al.~\cite{Damodaran_2015}} & Static features nearly as effective as dynamic features; \\
                                                                                               & hybrid approaches add complexity without consistent gain \\
\midrule
\multirow{2}{*}{Wen and Chow~\cite{WEN2021301128}} & Good accuracy using CNN models on fragmented inputs \\
                                                                                          & and effective for zero-day malware scenario \\
\bottomrule
\end{tabular}
}
\end{table}


Multimodal Machine Learning (MML) has developed as an innovative approach, utilizing diverse data modalities 
to enhance the performance of models. In~\cite{10041115}, Barua et al., review key advancements and 
challenges in MML, including the current state of the field, and they provide insights into 
challenges and \hbox{potential} solutions,
while also discussing several frameworks and trends. One significant finding in this paper is the growing trend toward 
deep learning approaches. These approaches have gained attention due to their ability to effectively process 
complex multimodal data, particularly in areas such as representation learning, fusion techniques, and alignment strategies. 
Several deep learning architectures are mentioned for their contributions to MML applications, 
including CNNs, Recurrent Neural Networks (RNN), and transformer-based architectures. 

In another paper, Zhai et al.~\cite{zhai2023odtc} proposed an online classification model for darknet traffic classification 
that integrates a CNN and a Bidirectional Gated Recurrent Unit (BiGRU) to extract spatial and temporal features from 
packet payloads. To enhance the overall representation of network traffic, the model incorporates flow-level abstract features 
processed by a Multilayer Perceptron (MLP). The authors state that this multimodal approach improves accuracy across 
multiple categories while reducing time and memory consumption by approximately~50\%. The authors claim that compared to 
state-of-the-art traffic classification models, the proposed method demonstrates superior classification performance.

Further, in~\cite{9478232}, Shobhit and Bera proposed a deep learning-based malware detection system with 
a unique architecture that combines image representations, text representations, and Generative Adversarial Networks (GANs). 
This so-called ModCGAN system architecture comprises several key components, including
an Image Autoencoder, a Text Autoencoder, and a GAN. They utilized log files containing API call 
sequences to generate both image and text representations. These representations are fed into 
the respective autoencoders and combined into a single feature vector. This feature vector is then 
used for malware classification purposes. The results demonstrated that this multi-modal approach 
effectively captures the dynamic behavior of malware, resulting in improved malware detection 
accuracy compared to traditional single-modality methods. 

Lastly, in~\cite{guo2023mdenetmultimodaldualembeddingnetworks}, Guo et al. propose another innovative approach 
called Multimodal Dual-Embedding Networks (MDENet), which they use to detect malware with highly similar features. 
MDENet leverages malware features from different modalities, such as malware images 
(generated from numeric features of malware samples) and malware ``sentences'' 
(created by organizing tokenized malware features into sentence-like structures). 
MDENet achieves good accuracy 
in both classifying known malware and detecting new, unknown families. Its effectiveness is validated through 
experiments on the popular Mailing dataset and the MAL-100+ dataset. This study provides further evidence
of the value of a multimodal approach in addressing challenging malware problems.

\section{Experiments and Results}\label{sect:exp}

In this section, we present experimental results comparing the performance of 
various multimodal approaches to typical machine learning models.
We train models using histograms, byte sequences, and bytes interpreted as 2-dimensional images,
depending on the model under consideration. For each of the three model types
discussed in Section~\ref{sect:ML_methods}, namely, SVM, LSTM, and CNN, 
we train a model on the header, we train another model on the sections, and we
train a third model on the entire file. Then we consider multimodal experiments,
where we train an SVM on the output of each of the nine combinations of 
header and section models.
We evaluate all of our models based on accuracy, which is computed as
\[
    \text{accuracy} = \frac{\text{TP} + \text{TN}}{\text{TP} + \text{TN} + \text{FP} + \text{FN}}
\]
where we have~$\text{TP} = \text{True Positives}$, $\text{TN} = \text{True Negatives}$, $\text{FP} = \text{False Positives}$, 
and~$\text{FN} = \text{False Negatives}$.

But, before we discuss our experimental results, 
we provide details on the dataset and features utilized in the experiments.

\subsection{Dataset}

All the experiments presented in this study are based on malware samples from the Malicia 
dataset---specifically, \texttt{release\_malicia\_1.0} dataset---which consists of~$11{,}371$ malicious 
files in raw PE file format, along with metadata that includes the class label 
for each sample~\cite{malicia-dataset}. 
The dataset contains a large number of families, many of which have only a few samples, 
so we restrict our attention to the following five malware families.

\begin{description}
\item[Cridex] is designed to steal banking credentials. In addition, after infecting a system, Cridex may integrate the compromised 
device into a botnet and exploit it for malicious activities, such as generating spam or executing 
Distributed Denial of Service (DDoS) attacks~\cite{cridex}.
\item[Harebot] is classified as both a backdoor and a rootkit. It installs itself on Windows systems without the user's consent, 
enabling attackers to gain unauthorized access to devices, networks, applications, and sensitive user information~\cite{harebot}.
\item[SecurityShield] is malware disguised as antivirus software. It secretly monitors and collects information 
about a user's activity while falsely reporting virus detections to coerce users into purchasing 
unnecessary software~\cite{securityshield}.
\item[Zbot]\!\!\!, also known as Zeus, is a Trojan that commonly spreads through spam emails, social media links, 
or drive-by downloads. It monitors infected systems to steal sensitive information such as banking credentials and 
passwords~\cite{zbot}.
\item[ZeroAccess] is a Trojan horse equipped with an advanced rootkit, allowing it to conceal its presence, 
create hidden files, install backdoors, and download additional malware for various malicious purposes.
\end{description}

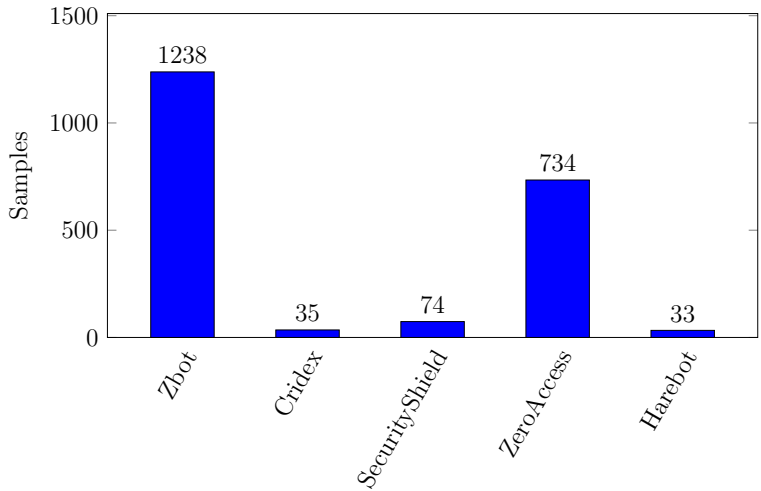
\begin{figure}[!htb]
\centering
\adjustbox{scale=0.8}{
\begin{tikzpicture}
\pgfkeys{/pgf/number format/.cd,1000 sep={}}
\begin{axis}[
    ybar=5*\pgflinewidth,
    bar width=30pt,
    symbolic x coords={Zbot, Cridex, SecurityShield, ZeroAccess, Harebot},
    xtick=data,
    nodes near coords,
    every node near coord/.append style={
    								scale=1.0,
								/pgf/number format/.cd,
								fixed,
								fixed zerofill,
								precision=0},
    ylabel={Samples},
    ymin=0,
    ymax=1510,
    ytick={0, 500, 1000, 1500},
    y tick label style={
    	scale=1.0,
	/pgf/number format/.cd,
   	fixed,
   	fixed zerofill,
	precision=0,
	},
    xtick=data,
    xtick style={color=white},
    x tick label style={scale=1.0,
        	rotate=60,
	anchor=north east,
	inner sep=0mm
	},
    width=0.8\textwidth,
    height=0.45\textwidth,
    enlarge x limits=0.15,
]
\addplot[fill=blue] coordinates {
(Zbot, 1238)
(Cridex, 35)
(SecurityShield, 74)
(ZeroAccess, 734)
(Harebot, 33)
};
\end{axis}
\end{tikzpicture}
}
\caption{Number of samples for each malware family}
\label{fig:malware_samples}
\end{figure}

Figure~\ref{fig:malware_samples} lists the number of samples that we consider 
from each of these malware families. Thus, the dataset for our experiments consists 
of~2114 samples and five classes, with a large imbalance between classes.

\subsection{SVM Results}

For the SVM model, we use byte histogram features. The raw byte sequences are extracted and interpreted as 
integer values ranging from~0 to~255. These integer values are then gathered into relative
histograms, which yields a feature vector of length~256, where each element is between~0 and~1,
and the sum of all~256 elements is~1, i.e., the vector satisfies the requirements of a 
discrete probability distribution. These histograms serve as the feature sets for our SVM classification models.

We use an~80-20 stratified split for training and testing and~5-fold cross-validation.
To determine the hyperparameters, we perform a grid search on the values
listed in Table~\ref{tab:SVM_hype}, using the header features. Note that the
values selected are in boldface in Table~\ref{tab:SVM_hype}.

\begin{table}[!htb]
\caption{SVM hyperparmaters tested (selected values in boldface)}\label{tab:SVM_hype}
\centering
\adjustbox{scale=0.85}{
\begin{tabular}{c|ll}
\toprule
\textbf{Hyperparameter} & \ \ \ \textbf{Values tested} & \ \ \ \textbf{Description} \\ \midrule
        $C$ & \{0.001, 0.01, 0.1, 1, \textbf{10}\} & \text{Regularization parameter} \\
        $\gamma$ & \{0.001, \textbf{0.01}, 0.1, 1\} & \text{Kernel coefficient} \\
        \text{kernel} & \{\text{poly}, \text{linear}, \textbf{rbf}\} & \text{Kernel type}\\ 
\bottomrule
\end{tabular}
}
\end{table}

\begin{figure}[!htb]
    \centering
    \adjustbox{scale=0.8}{
    \begin{tikzpicture}
\pgfkeys{/pgf/number format/.cd,1000 sep={}}
\begin{axis}[
    ybar=5*\pgflinewidth,
    bar width=40pt,
    symbolic x coords={Header, Sections, Entire file},
    xtick=data,
    nodes near coords,
    every node near coord/.append style={
    								scale=1.0,
								/pgf/number format/.cd,
								fixed,
								fixed zerofill,
								precision=4},
    ylabel={Accuracy},
    ymin=0.5,
    ymax=1.05,
    ytick={0.50, 0.60, 0.70, 0.80, 0.90, 1.00},
    y tick label style={
    	scale=1.0,
	/pgf/number format/.cd,
   	fixed,
   	fixed zerofill,
	precision=2,
	},
    xtick=data,
    xtick style={color=white},
    x tick label style={scale=1.0,
	},
    width=0.65\textwidth,
    height=0.45\textwidth,
    enlarge x limits=0.225,
]
\addplot[fill=blue] coordinates {
(Header, 0.9606)
(Sections, 0.9835)
(Entire file, 0.9811)
};
\end{axis}
\end{tikzpicture}
    }
    \caption{SVM: Accuracy comparison across experiments}
    \label{fig:accuracy-comparison-SVM}
\end{figure}
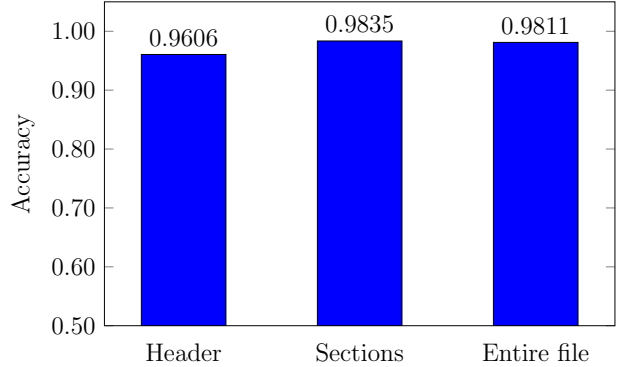

In Figure~\ref{fig:accuracy-comparison-SVM}, we observe that the SVM performs well 
when applied to histogram features. The best results are for a model is trained on only 
the sections, while the header model performs slightly worse and the entire
file gives almost the same results as the sections model.

%
%
%

\subsection{LSTM Results}

Our LSTM models are trained on byte sequences. Since the PE header is typically~324 bytes, this is the 
sequence length used for our header model. For efficiency, when training the sections model and the ``entire'' PE file model, 
we experiment with lengths of~1000 and~2000 bytes. 

The structure of the LSTM models is detailed in 
Table~\ref{tab:lstm-architecture}. Note that the only variation between the models
is the input sequence length. 

\begin{table}[!htb]
    \caption{Hyperparameters and architecture for LSTM}\label{tab:lstm-architecture}
    \centering
    \adjustbox{scale=0.85}{
    \begin{tabular}{c|c}
    \toprule
    \textbf{Hyperparameter}                   & \textbf{Value} \\ \midrule
    Batch size                           & 30             \\
    Epochs             & 10             \\
    Data used for validation & 20\%           \\
    Input sequence length (bytes)        & 324            \\
    Number of LSTM units                 & 64             \\
    Activation function (hidden layer)   & Sigmoid        \\
    Number of dense units (output layer) & 5              \\
    Activation function (output layer)   & Softmax        \\
    Loss function                        & Categorical cross-entropy \\
    Optimizer                            & Adam           \\ \bottomrule
    \end{tabular}
    }
\end{table}

\begin{figure}[!htb]
    \centering
    \adjustbox{scale=0.8}{
    \begin{tikzpicture}
\pgfkeys{/pgf/number format/.cd,1000 sep={}}
\begin{axis}[
    ybar=5*\pgflinewidth,
    bar width=30pt,
    symbolic x coords={Header, Sections (1000), Sections (2000), Entire file (1000), Entire file (2000)},
    xtick=data,
    nodes near coords,
    every node near coord/.append style={
    								scale=1.0,
								/pgf/number format/.cd,
								fixed,
								fixed zerofill,
								precision=4},
    ylabel={Accuracy},
    ymin=0.5,
    ymax=1.05,
    ytick={0.50, 0.60, 0.70, 0.80, 0.90, 1.00},
    y tick label style={
    	scale=1.0,
	/pgf/number format/.cd,
   	fixed,
   	fixed zerofill,
	precision=2,
	},
    xtick=data,
    xtick style={color=white},
    x tick label style={scale=1.0,
        	rotate=60,
	anchor=north east,
	inner sep=0mm
	},
    width=0.8\textwidth,
    height=0.45\textwidth,
    enlarge x limits=0.175,
]
            \addplot[fill=blue, draw=none] coordinates {
                (Header, 0.8865)
                (Sections (1000), 0.5794)
                (Sections (2000), 0.6031)
                (Entire file (1000), 0.6024)
                (Entire file (2000), 0.5800)
            };
        \end{axis}
    \end{tikzpicture}
    }
    \caption{LSTM: Accuracy comparison across experiments}
    \label{fig:accuracy-comparison-LSTM}
\end{figure}
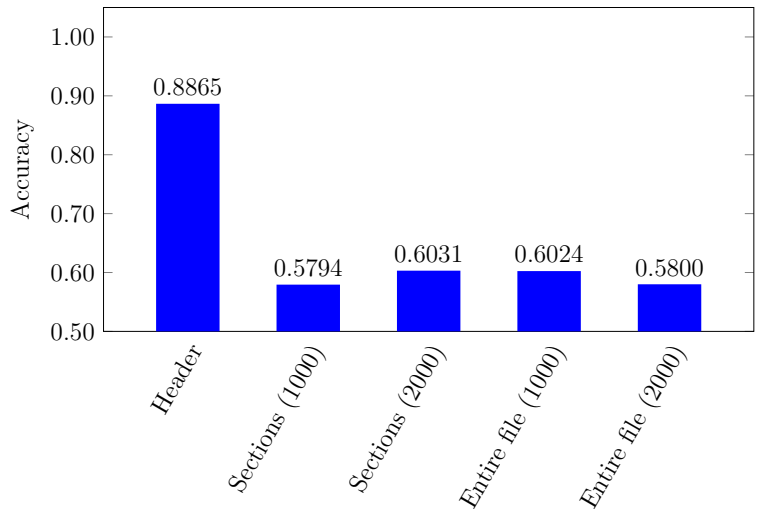

From Figure~\ref{fig:accuracy-comparison-LSTM} we observe that the LSTM model performed reasonably well on 
when trained in the PE header, the other models yielded very poor results. The malware files under consideration
are typically much larger than~2000 bytes, and hence better results might be obtainable using longer sequence
lengths. However, LSTM models are computationally expensive to train for large sequence lengths,
and we have already obtained strong results for the sections and entire file cases using SVM models.

\subsection{CNN Results}

For our CNN models, we convert the raw bytes into simple images as follows. 
For the header model, we generate a~$16\times 16$ ``byte image'' from the first~256
bytes of the PE header as
$$
  \mbox{header image} = 
  \left[
  \begin{array}{ccccc}
  b_{0} & b_{1} & b_{2} & \cdots & b_{15} \\
  b_{16} & b_{17} & b_{18} & \cdots & b_{31} \\
  b_{32} & b_{33} & b_{34} & \cdots &  b_{47} \\
  \vdots & \vdots & \vdots & \ddots & \vdots \\
  b_{240} & b_{241} & b_{242} & \cdots & b_{255}
  \end{array}
  \right]
$$
where~$b_{k}$ is the~$k^{\thth}$ byte of the PE file. In an analogous manner we construct 
byte images of size~$32\times 32$ for both the sections and entire file features, beginning at
byte~$b_{324}$ in the sections case, and beginning at byte~$b_{0}$ for the entire file case.
Note that in both of these latter two cases, we are considering precisely~1024 bytes.
These byte images are then converted into grayscale images using the Python library Pillow~\cite{image_lib}. 
The conversion process involves interpreting the binary data as a sequence of 
unsigned 8-bit integers, and each value corresponds to the intensity of the corresponding 
grayscale pixel, and the pixel intensity values are then assigned to the image using the 
\texttt{putdata} method. 

We experimented with various CNN architectures.
Details on the selected architecture are given in Table~\ref{tab:cnn-architecture}.
\begin{table}[!htb]
    \caption{CNN model details}\label{tab:cnn-architecture}
    \centering
    \adjustbox{scale=0.85}{
    \begin{tabular}{c|cc}
    \toprule
    \multirow{2}{*}{\textbf{Layer (type)}} & \multirow{2}{*}{\textbf{Output shape}} & \textbf{Number of} \\
            &     & \textbf{parameters} \\ \midrule
    \texttt{conv2d (Conv2D)} & (None, 30, 30, 32) & 320 \\
    \texttt{max\_pooling2d (MaxPooling2D)} & (None, 15, 15, 32) & 0 \\
    \texttt{flatten (Flatten)} & (None, 7200) & 0 \\
    \texttt{dense\_1 (Dense)} & (None, 128) & 921,728 \\
    \texttt{dense\_2 (Dense)} & (None, 5) & 645 \\ \bottomrule
    \end{tabular}
    }
\end{table}

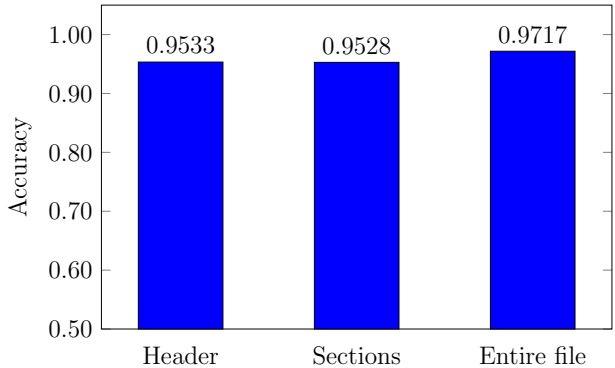
\begin{figure}[!htb]
    \centering
    \adjustbox{scale=0.8}{
    \begin{tikzpicture}
\pgfkeys{/pgf/number format/.cd,1000 sep={}}
\begin{axis}[
    ybar=5*\pgflinewidth,
    bar width=40pt,
    symbolic x coords={Header, Sections, Entire file},
    xtick=data,
    nodes near coords,
    every node near coord/.append style={
    								scale=1.0,
								/pgf/number format/.cd,
								fixed,
								fixed zerofill,
								precision=4},
    ylabel={Accuracy},
    ymin=0.5,
    ymax=1.05,
    ytick={0.50, 0.60, 0.70, 0.80, 0.90, 1.00},
    y tick label style={
    	scale=1.0,
	/pgf/number format/.cd,
   	fixed,
   	fixed zerofill,
	precision=2,
	},
    xtick=data,
    xtick style={color=white},
    x tick label style={scale=1.0,
	},
    width=0.65\textwidth,
    height=0.45\textwidth,
    enlarge x limits=0.225,
]
            \addplot[fill=blue] coordinates {
                (Header, 0.9533)
                (Sections, 0.9528)
                (Entire file, 0.9717)
            };
        \end{axis}
    \end{tikzpicture}
    }
    \caption{CNN: Accuracy comparison across experiments}
    \label{fig:accuracy-comparison-CNN}
\end{figure}

Figure~\ref{fig:accuracy-comparison-CNN} shows that our CNN models perform well in all three cases,
with images based on the first~1024 bytes of the entire file yielding the best accuracy. 
These results demonstrates the efficacy of image-based techniques in general---and CNNs in particular---in 
the realm of malware classification.

\subsection{Multimodal Models}\label{sect:MM}

For our multimodal experiments, we test all combinations of one of our header-based models with 
one of our sections-based models. Since we trained models with header features
and models with sections features for each of three distinct cases (SVM, LSTM, and CNN), 
we have a total of nine multimodal combinations.

To combine our models in a multimodal fashion, the output
layer probabilities of the header-based model and the sections-based model are
concatenated, and the resulting vector serves as the feature vector for an SVM classifier.
Since each model is trained to distinguish between five classes, the feature vectors
for the final SVM classifier are all of length~10. 

We note in passing that for this
multimodal approach, training the header-based model and the sections-based model
can be viewed as feature engineering steps, since these component models
take the raw features as input and produce new features for the final SVM classifier.
We also note that this feature engineering step is straightforward for LSTM and CNN models, 
but an SVM does not directly generate probabilities. Therefore, for our header-based
and sections-based SVM models, we set \texttt{svm.SVC(probability=True)} in \texttt{scikit-learn}, 
which introduces a post-training probability calculation step.

The results for all of our multimodal models are summarized in Figure~\ref{fig:mult} 
of the Appendix.
In these results, multimodal models are specified in the form
$$
  (M_h,M_s)\rightarrow M_m
$$
where~$M_h$ is the model trained using header-based features, 
$M_s$ is the model trained using section-based features,
and~$M_m$ is the multimodal model that acts as the classifier, based on the output---in the
form of vectors of probabilities---of the models~$M_h$ and~$M_s$.
Each subfigure represents the accuracy---given to two decimal 
places---achieved by different combinations of models. Note that
Figures~\ref{fig:mult}(a) through~(c) are sorted by the header model,
while Figures~\ref{fig:mult}(d) through~(f) give the same results, but
sorted by the sections model. As mentioned above, in all cases, SVM is used as the
multimodal~$M_m$ classifier.

%

\begin{figure}[!htb]
    \centering
    \adjustbox{scale=0.8}{
    \begin{tikzpicture}
\pgfkeys{/pgf/number format/.cd,1000 sep={}}
\begin{axis}[
    ybar=5*\pgflinewidth,
    bar width=40pt,
    symbolic x coords={A, B, C, D},
    xticklabels={SVM (sections), LSTM (header), CNN (entire file), Multimodal},
    xtick=data,
    nodes near coords,
    every node near coord/.append style={
    								scale=1.0,
								/pgf/number format/.cd,
								fixed,
								fixed zerofill,
								precision=4},
    ylabel={Accuracy},
    ymin=0.5,
    ymax=1.055,
    ytick={0.50, 0.60, 0.70, 0.80, 0.90, 1.00},
    y tick label style={
    	scale=1.0,
	/pgf/number format/.cd,
   	fixed,
   	fixed zerofill,
	precision=2,
	},
    xtick=data,
    xtick style={color=white},
    x tick label style={scale=1.0,
        	rotate=60,
	anchor=north east,
	inner sep=0mm
	},
    width=0.75\textwidth,
    height=0.45\textwidth,
    enlarge x limits=0.225,
]
            \addplot[fill=blue, draw=none] coordinates {
                (A, 0.9835)
                (B, 0.8865)
                (C, 0.9717)
                (D, 0.9930)
            };
        \end{axis}
    \end{tikzpicture}
    }
    \caption{Accuracy comparison across all experiments}
    \label{fig:accuracy-comparison-multimodal}
\end{figure}
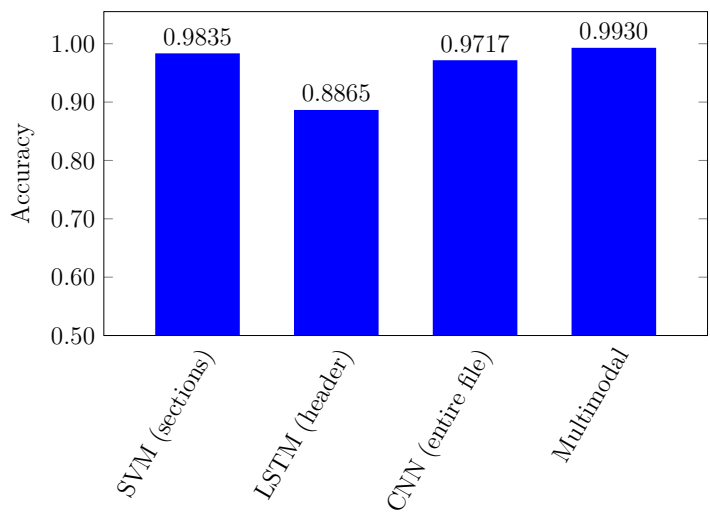

Figure~\ref{fig:accuracy-comparison-multimodal} provides a comparison of 
the best multimodal model result to the best results obtained
with each of the individual models, namely, SVM, LSTM, and CNN.
The best of the multimodal models are
$$
  (\mbox{LSTM},\mbox{CNN})\rightarrow \mbox{SVM} \mbox{\ \ and\ \ }
  (\mbox{CNN},\mbox{CNN})\rightarrow \mbox{SVM}
$$
which achieve accuracies of~0.9929 and~0.9930, respectively.
We observe that these best multimodal accuracies
are about~1\%\ better than the best of the individual models. 

While a~1\%\ improvement may not seem overly 
impressive at first glance, it is important to note 
that the second-best model had already attained more than~98\%\ accuracy.
Therefore, the results show that the number of samples misclassified
by our best multimodal model is about half the number that were
misclassified by the best of the individual models. Thus, we conclude that
the multimodal approach to PE malware classification considered in this paper
appears to have genuine merit.

\section{Conclusion and Future Work}\label{sect:con}

In this paper, we demonstrated the effectiveness of combining multiple machine-learning models 
in a multimodal sense can enhance malware classification accuracy. 
For our baseline cases, we trained SVM, LSTM, and CNN models on features extracted from the
PE header, and on features extracted from the other sections of the PE file, and we also trained
these same models on the overall PE file. We then generated multimodal models, where the output layer
probabilities of a PE-header model and a PE-sections model were combined and used as feature vectors 
to train an SVM classifier. In this multimodal approach, we were able to take advantage of the unique strengths 
of specific learning techniques on different parts of the Windows PE file, which
improved classification accuracy, as compared to the baseline cases.

There are many potential directions for improvements and extensions of the research presented in this paper. 
Testing the techniques considered in this paper on a larger, more diverse, and more challenging
dataset would aid in more accurately quantifying the effectiveness of such a multimodal approach.
We could also consider additional features, such as opcode sequences, API calls, and various
graph structures, and we could employ advanced feature engineering techniques, 
such as Word2Vec~\cite{Chandak2021}. In a similar vein, 
it has been demonstrated that the method used to generate images from raw data
can affect the accuracy of CNN classifiers~\cite{Rust}, and hence more sophisticated
image-generation techniques should be considered.
Experiments involving additional learning techniques would also be interesting---in particular, 
Graph Neural Networks have shown considerable promise in the malware domain~\cite{vrinda},
and such techniques are likely to provide unique strengths in a multimodal setting.

\bibliographystyle{plain}
\bibliography{references}

\section*{Appendix}\label{app:a}

\titleformat{\section}{\normalfont\large\bfseries}{}{0em}{#1\ \thesection}
\setcounter{section}{0}
\renewcommand{\thesection}{\Alph{section}}
\renewcommand{\thesubsection}{A.\arabic{subsection}}
\setcounter{table}{0}
\renewcommand{\thetable}{A.\arabic{table}}
\setcounter{figure}{0}
\renewcommand{\thefigure}{A.\arabic{figure}}

The results of all multimodal combinations tested 
are given in Figure~\ref{fig:mult}. For additional details
on the multimodal experiments used to generate these
results, see Section~\ref{sect:MM}.

\begin{figure}[!htb]
    \centering
    \begin{tabular}{cc}
    \adjustbox{scale=0.65}{
    \begin{tikzpicture}
\pgfkeys{/pgf/number format/.cd,1000 sep={}}
\begin{axis}[
    ybar=5*\pgflinewidth,
    bar width=40pt,
    symbolic x coords={A,B,C},
    xticklabels={$(\mbox{SVM,\,SVM})\rightarrow \mbox{SVM}$,
    		       $(\mbox{SVM,\,LSTM})\rightarrow \mbox{SVM}$, 
		       $(\mbox{SVM,\,CNN})\rightarrow \mbox{SVM}$},
    xtick=data,
    nodes near coords,
    every node near coord/.append style={
    								scale=1.0,
								/pgf/number format/.cd,
								fixed,
								fixed zerofill,
								precision=4},
    ylabel={Accuracy},
    ymin=0.5,
    ymax=1.05,
    ytick={0.50, 0.60, 0.70, 0.80, 0.90, 1.00},
    y tick label style={scale=1.0,
	/pgf/number format/.cd,
   	fixed,
   	fixed zerofill,
	precision=2,
	},
    xtick=data,
    xtick style={color=white},
    x tick label style={scale=0.9,
        	rotate=60,
	anchor=north east,
	inner sep=0mm
	},
    width=0.65\textwidth,
    height=0.45\textwidth,
    enlarge x limits=0.225,
]
\addplot[fill=blue, draw=none] coordinates {
(A, 0.5933)
(B, 0.7966)
(C, 0.9434)
};
\end{axis}
\end{tikzpicture}
    }
    & 
    \adjustbox{scale=0.65}{
    \begin{tikzpicture}
\pgfkeys{/pgf/number format/.cd,1000 sep={}}
\begin{axis}[
    ybar=5*\pgflinewidth,
    bar width=40pt,
    symbolic x coords={A,B,C},
    xticklabels={$(\mbox{LSTM,\,SVM})\rightarrow \mbox{SVM}$,
    		       $(\mbox{LSTM,\,LSTM})\rightarrow \mbox{SVM}$, 
		       $(\mbox{LSTM,\,CNN})\rightarrow \mbox{SVM}$},
    xtick=data,
    nodes near coords,
    every node near coord/.append style={
    								scale=1.0,
								/pgf/number format/.cd,
								fixed,
								fixed zerofill,
								precision=4},
    ylabel={Accuracy},
    ymin=0.5,
    ymax=1.05,
    ytick={0.50, 0.60, 0.70, 0.80, 0.90, 1.00},
    y tick label style={
    	scale=1.0,
	/pgf/number format/.cd,
   	fixed,
   	fixed zerofill,
	precision=2,
	},
    xtick=data,
    xtick style={color=white},
    x tick label style={scale=0.9,
        	rotate=60,
	anchor=north east,
	inner sep=0mm
	},
    width=0.65\textwidth,
    height=0.45\textwidth,
    enlarge x limits=0.225,
]
            \addplot[fill=blue, draw=none] coordinates {
                (A, 0.9125)
                (B, 0.9361)
                (C, 0.9929)
            };
        \end{axis}
    \end{tikzpicture}
    }
    \\
    \adjustbox{scale=0.85}{(a) SVM header models}
    & 
    \adjustbox{scale=0.85}{(b) LSTM header models}
    \\ \\[-1.25ex]
    \adjustbox{scale=0.65}{
    \begin{tikzpicture}
\pgfkeys{/pgf/number format/.cd,1000 sep={}}
\begin{axis}[
    ybar=5*\pgflinewidth,
    bar width=40pt,
    symbolic x coords={A,B,C},
    xticklabels={$(\mbox{CNN,\,SVM})\rightarrow \mbox{SVM}$,
    		       $(\mbox{CNN,\,LSTM})\rightarrow \mbox{SVM}$, 
		       $(\mbox{CNN,\,CNN})\rightarrow \mbox{SVM}$},
    xtick=data,
    nodes near coords,
    every node near coord/.append style={
    								scale=1.0,
								/pgf/number format/.cd,
								fixed,
								fixed zerofill,
								precision=4},
    ylabel={Accuracy},
    ymin=0.5,
    ymax=1.05,
    ytick={0.50, 0.60, 0.70, 0.80, 0.90, 1.00},
    y tick label style={
    	scale=1.0,
	/pgf/number format/.cd,
   	fixed,
   	fixed zerofill,
	precision=2,
	},
    xtick=data,
    xtick style={color=white},
    x tick label style={scale=0.9,
        	rotate=60,
	anchor=north east,
	inner sep=0mm
	},
    width=0.65\textwidth,
    height=0.45\textwidth,
    enlarge x limits=0.225,
]
            \addplot[fill=blue, draw=none] coordinates {
                (A, 0.9514)
                (B, 0.9621)
                (C, 0.9930)
            };
        \end{axis}
    \end{tikzpicture}
    }
    & 
    \adjustbox{scale=0.65}{
    \begin{tikzpicture}
\pgfkeys{/pgf/number format/.cd,1000 sep={}}
\begin{axis}[
    ybar=5*\pgflinewidth,
    bar width=40pt,
    symbolic x coords={A,B,C},
    xticklabels={$(\mbox{SVM,\,SVM})\rightarrow \mbox{SVM}$,
    		       $(\mbox{LSTM,\,SVM})\rightarrow \mbox{SVM}$, 
		       $(\mbox{CNN,\,SVM})\rightarrow \mbox{SVM}$},
    xtick=data,
    nodes near coords,
    every node near coord/.append style={
    								scale=1.0,
								/pgf/number format/.cd,
								fixed,
								fixed zerofill,
								precision=4},
    ylabel={Accuracy},
    ymin=0.5,
    ymax=1.05,
    ytick={0.50, 0.60, 0.70, 0.80, 0.90, 1.00},
    y tick label style={scale=1.0,
	/pgf/number format/.cd,
   	fixed,
   	fixed zerofill,
	precision=2,
	},
    xtick=data,
    xtick style={color=white},
    x tick label style={scale=0.9,
        	rotate=60,
	anchor=north east,
	inner sep=0mm
	},
    width=0.65\textwidth,
    height=0.45\textwidth,
    enlarge x limits=0.225,
]
\addplot[fill=red, draw=none] coordinates {
(A, 0.5933)
(B, 0.9125)
(C, 0.9514)
};
\end{axis}
\end{tikzpicture}
    }
    \\
    \adjustbox{scale=0.85}{(c) CNN header models}
    & 
    \adjustbox{scale=0.85}{(d) SVM sections models}
    \\ \\[-1.25ex]
    \adjustbox{scale=0.65}{
    \begin{tikzpicture}
\pgfkeys{/pgf/number format/.cd,1000 sep={}}
\begin{axis}[
    ybar=5*\pgflinewidth,
    bar width=40pt,
    symbolic x coords={A,B,C},
    xticklabels={$(\mbox{SVM,\,LSTM})\rightarrow \mbox{SVM}$,
    		       $(\mbox{LSTM,\,LSTM})\rightarrow \mbox{SVM}$, 
		       $(\mbox{CNN,\,LSTM})\rightarrow \mbox{SVM}$},
    xtick=data,
    nodes near coords,
    every node near coord/.append style={
    								scale=1.0,
								/pgf/number format/.cd,
								fixed,
								fixed zerofill,
								precision=4},
    ylabel={Accuracy},
    ymin=0.5,
    ymax=1.05,
    ytick={0.50, 0.60, 0.70, 0.80, 0.90, 1.00},
    y tick label style={
    	scale=1.0,
	/pgf/number format/.cd,
   	fixed,
   	fixed zerofill,
	precision=2,
	},
    xtick=data,
    xtick style={color=white},
    x tick label style={scale=0.9,
        	rotate=60,
	anchor=north east,
	inner sep=0mm
	},
    width=0.65\textwidth,
    height=0.45\textwidth,
    enlarge x limits=0.225,
]
            \addplot[fill=red, draw=none] coordinates {
                (A, 0.7966)
                (B, 0.9361)
                (C, 0.9621)
            };
        \end{axis}
    \end{tikzpicture}
    }
    & 
    \adjustbox{scale=0.65}{
    \begin{tikzpicture}
\pgfkeys{/pgf/number format/.cd,1000 sep={}}
\begin{axis}[
    ybar=5*\pgflinewidth,
    bar width=40pt,
    symbolic x coords={A,B,C},
    xticklabels={$(\mbox{SVM,\,CNN})\rightarrow \mbox{SVM}$,
    		       $(\mbox{LSTM,\,CNN})\rightarrow \mbox{SVM}$, 
		       $(\mbox{CNN,\,CNN})\rightarrow \mbox{SVM}$},
    xtick=data,
    nodes near coords,
    every node near coord/.append style={
    								scale=1.0,
								/pgf/number format/.cd,
								fixed,
								fixed zerofill,
								precision=4},
    ylabel={Accuracy},
    ymin=0.5,
    ymax=1.05,
    ytick={0.50, 0.60, 0.70, 0.80, 0.90, 1.00},
    y tick label style={
    	scale=1.0,
	/pgf/number format/.cd,
   	fixed,
   	fixed zerofill,
	precision=2,
	},
    xtick=data,
    xtick style={color=white},
    x tick label style={scale=0.9,
        	rotate=60,
	anchor=north east,
	inner sep=0mm
	},
    width=0.65\textwidth,
    height=0.45\textwidth,
    enlarge x limits=0.225,
]
            \addplot[fill=red, draw=none] coordinates {
                (A, 0.9434)
                (B, 0.9920)
                (C, 0.9930)
            };
        \end{axis}
    \end{tikzpicture}
    }
    \\
    \adjustbox{scale=0.85}{(e) LSTM sections models}
    & 
    \adjustbox{scale=0.85}{(f) CNN sections models}
    \\ 
    \end{tabular}
    \caption{Accuracy of multimodal combinations}\label{fig:mult}
\end{figure}

\end{document}